\documentclass[sigconf]{acmart}

\usepackage{subcaption}
\usepackage{graphicx}
\usepackage{booktabs}
\usepackage{hyperref}
\usepackage[capitalise,nameinlink,noabbrev]{cleveref}

\usepackage{xcolor}

\usepackage[font=small,labelfont=bf]{caption}
\AtBeginDocument{%
  }

\setcopyright{none}
\copyrightyear{2025}
\acmYear{2025}
\acmDOI{10.1145/xxxxxxx.xxxxxxx}
\acmISBN{978-x-xxxx-xxxx-x/25/10}
\acmConference[CAMS'25]{Computer Architecture Modeling and Simulation}{October, 2025}{Seoul, Republic of Korea}
\acmBooktitle{The 3rd Workshop on Computer Architecture Modeling and Simulation
(CAMS 2025), Oct 18, 2025, Seoul, Korea}
\acmISBN{978-1-4503-XXXX-X/2018/06}

\acmSubmissionID{4898}

\usepackage{xspace}

\def\arch{gem5 Co-Pilot\xspace}

\begin{document}

\title{gem5 Co-Pilot: AI Assistant Agent for Architectural Design Space Exploration}


\author{Zuoming Fu}
\authornote{These authors contributed equally to this research.}
\affiliation{
  \institution{Cornell University}
  \city{Ithaca}
  \state{NY}
  \country{USA}
}

\author{Alex Manley}
\authornotemark[1]
\affiliation{
  \institution{University of Kansas}
  \city{Lawrence}
  \state{KS}
  \country{USA}
}

 \author{Mohammad Alian}
 \authornote{Corresponding author}
 \affiliation{
   \institution{Cornell University}
   \city{Ithaca}
   \state{NY}
   \country{USA}
 }

% \affiliation{%

% \affiliation{%

% \affiliation{%

% \affiliation{%

% \affiliation{%

% \affiliation{%

% \affiliation{%

\renewcommand\footnotetextcopyrightpermission[1]{} % removes permission footer

\begin{abstract}
Generative AI is increasing the productivity of software and hardware development across many application domains. In this work, we utilize the power of Large Language Models (LLMs) to develop a co-pilot agent for assisting gem5 users with automating design space exploration. Computer architecture design space exploration is complex and time-consuming, given that numerous parameter settings and simulation statistics must be analyzed before improving the current design. The emergence of LLMs has significantly accelerated the analysis of long-text data as well as smart decision-making, two key functions in a successful design space exploration task. In this project, we first build \arch, an AI agent assistant for gem5, which comes with a webpage-GUI for smooth user interaction, agent automation, and result summarization. We also implemented a language for design space exploration, as well as a Design Space Database (DSDB). With DSDB, \arch effectively implements a Retrieval Augmented Generation system for gem5 design space exploration. We experiment on cost-constraint optimization with four cost ranges and compare our results with two baseline models. Results show that \arch can quickly identify optimal parameters for specific design constraints based on performance and cost, with limited user interaction.

%This article focuses on using Large Language Model (LLM) to accelerate the process of computer architecture (CA) design space computer exploration (DSE). The complexity and time-consumption of CA DSE have been extremely high given that tons of parameter settings and simulation stats must be analyzed before proceeding to improve the current design. The emergence of LLMs has astonishingly accelerated the analysis of long-text data as well as smart decision-making, which are two key functions in a successful DSE task. In this project, we built an AI Agent embedded with gem5, ariad DSL, LLM, and  streamlit UI for accelerating the DSE task. Results show that our program can run DSE automatically to find the top 5\% best parameter set at a cost of around \$0.5 per task.

\end{abstract}





\maketitle

\section{Introduction}

Computer architecture design space exploration (DSE) is the process of finding optimal parameters in a design space given a goal and constraints. In this project, we define a \textit{design space} by its goals, parameters, and constraints, and DSE as the process (by human, machine, or both) of identifying the best parameters for a goal under constraints. Computer architecture DSE is simply this process instantiated in architecture design.

A typical example of DSE is exploring the cache hierarchy design space of a CPU microarchitecture to maximize performance under certain constraints (such as power and/or area) for a specific workload (for example, a matrix–matrix multiplication kernel).

Traditional human-driven DSE is tedious: architects must explore thousands of parameter values, run tens to hundreds of simulations, and analyze large volumes of statistics to understand the effects of different parameters. LLMs offer a way to accelerate this labor-intensive process. With their long-context capabilities, LLMs can process simulation data and scripts, provide well-reasoned decisions or suggestions, and, when paired with function-calling, act as autonomous agents that invoke external tools for more effective decision-making.  

In this work, we present \arch, an LLM-powered assistant for computer architecture DSE using gem5~\cite{lowe2020gem5,binkert2011gem5}. To evaluate \arch, we explore the design space of an L2 cache under a total power constraint for a workload. Results show that \arch can reach near-optimal configurations within 1--8 generation stages (see \cref{state machine}) and 2--12 gem5 runs. Using GPT-4o via OpenAI’s cloud API, each DSE session for a given constraint costs less than \$0.5.

%Python, and Streamlit~\cite{streamlit}. It supports arbitrary design tasks and workloads. To evaluate \arch, we study an L2 cache design space under constraints such as total power, area, and EDP, focusing on total power in our experiments while leaving others supported. Results show that \arch can reach near-optimal configurations within 1–8 GEN stages (see \cref{state machine}) and 2–12 gem5 runs. Using GPT-4o via OpenAI’s cloud, each DSE for a given constraint costs under \$0.5.

\section{Background}

Computer architecture Design Space Exploration (DSE) aims to identify optimal design configurations under constraints such as performance, power, area, and cost. The vast and complex nature of the design space makes exhaustive evaluation impractical, driving the need for efficient and scalable exploration strategies.

Traditional DSE methods include brute-force search, parameter sweeps, and heuristic algorithms. Brute-force guarantees optimality but does not scale with problem size~\cite{dse-traditional}. Heuristics such as simulated annealing and particle swarm optimization~\cite{simulated-annealing} focus search on promising regions, though they may converge to local optima and often require careful hyperparameter tuning. Grid search 
% \malian{add citation}
covers the parameter space systematically but suffers from exponential growth \cite{Li_ICS04}. Architectural DSE often relies on slow cycle-accurate simulators like gem5, which further limits the practicality of these approaches. Domain-specific tools such as BOOM-Explorer~\cite{Arch-2021ICCAD-BOOM-Explorer} improve coverage but remain constrained by simulator overhead.

Machine learning offers more scalable alternatives by replacing simulations with predictive models. Techniques such as reinforcement learning (RL)~\cite{rl-dse}, genetic algorithms~\cite{genetic-dse}, ant colony optimization~\cite{aco-dse}, and Bayesian optimization (BO)~\cite{bayesian-dse} have been successfully applied. RL formulates DSE as a sequential decision process, adapting efficiently across workloads. Evolutionary and swarm intelligence methods enable efficient global exploration, while BO uses probabilistic surrogates for sample-efficient optimization. Frameworks like ArchGym~\cite{archgym} integrate multiple ML agents for trade-off optimization, and tools such as ZigZag~\cite{Mei2021ZigZag} employ analytical modeling to avoid slow simulations for DNN accelerators. More recently, Pareto-driven active learning~\cite{Zhai2023ParetoActiveLearning} combines active learning with Pareto optimality to accelerate multi-objective DSE.

Despite these advances, ML-based DSE still faces challenges, including high training cost, low volumes of training data, limited generalization, and interpretability issues. Real-world applications require not only efficiency but also explainability. The emergence of generative large language models (LLMs) offers a promising direction for incorporating pre-trained reasoning into DSE workflows. Future efforts will likely focus on hybrid strategies (combining RL, BO, and heuristics) while incorporating additional constraints such as power, thermal, and security into the optimization process.

\section{System Design}

\begin{figure}
\centering
\includegraphics[width=0.8\linewidth]{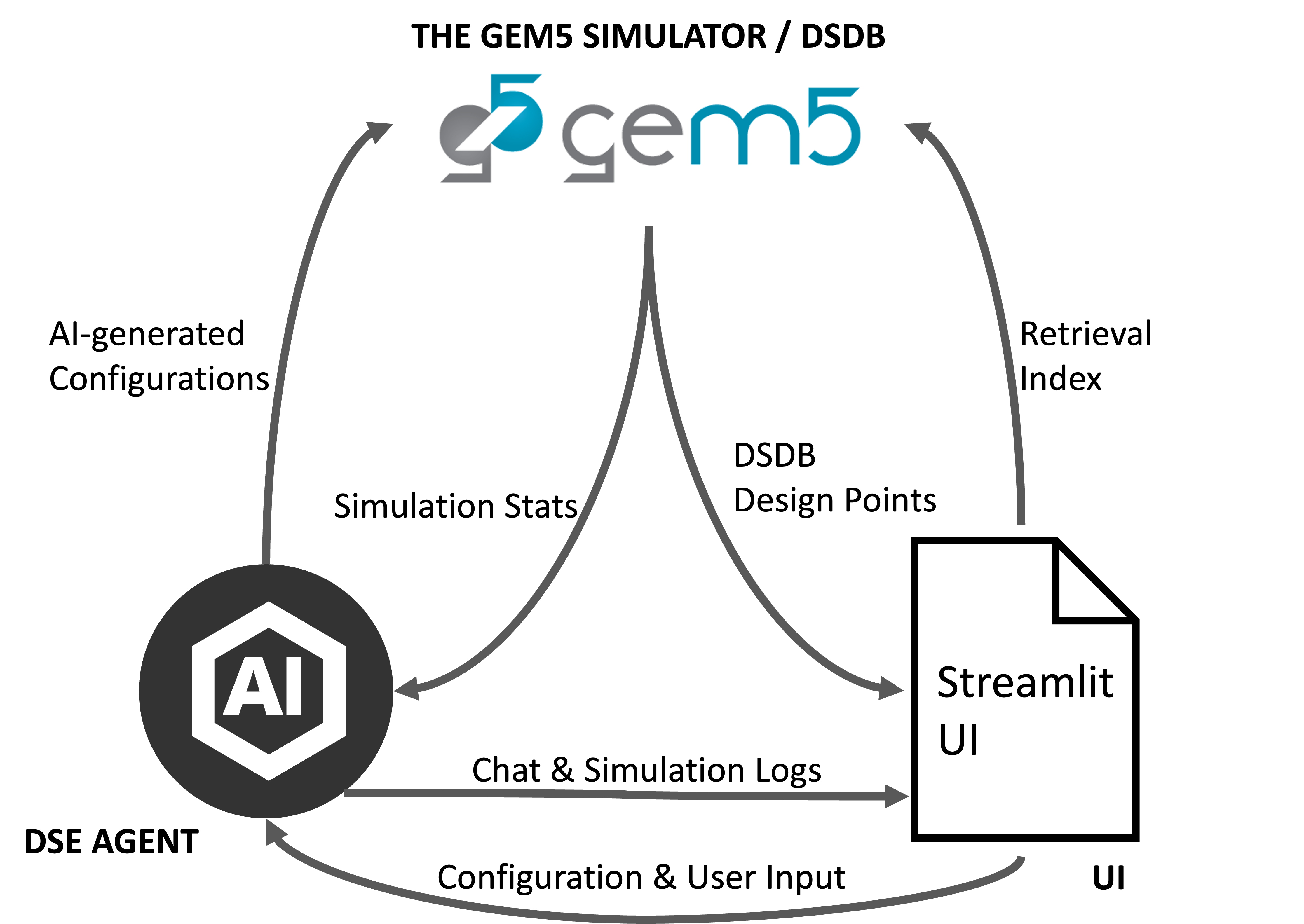}
\caption{DSE system overview}
\label{DSE system overview}
\end{figure}

Our DSE system comprises three core components: a DSE AI agent, the gem5 simulator (or DSDB, see \cref{DSDB}), and a Streamlit-based user interface (UI), as illustrated in \cref{DSE system overview}. The architecture separates the agent, simulation unit, and UI to improve flexibility and extensibility. The DSE agent, built around an LLM-driven state machine, dispatches gem5 configuration parameters to the simulator and relays contextual data (e.g., chat history) and simulation logs to the UI. The gem5 simulator (or DSDB) returns simulation results to the agent for optimization. The Streamlit UI retrieves design points from the DSDB for visualization and manages system configuration and user inputs for the agent.

\begin{figure}
\centering
\includegraphics[width=0.8\linewidth]{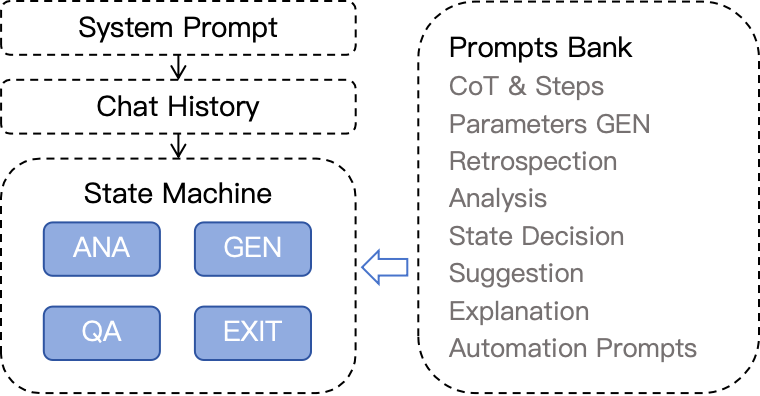}
\caption{DSE agent Structure}
\label{fig:agent}
\end{figure}

As depicted in \cref{fig:agent}, the AI agent acts as the central controller. A system prompt defines high-level goals, including DSE rules and permitted states, to steer the LLM’s operation. This prompt, combined with chat history, provides context for each LLM invocation. A state machine with four states (\texttt{ANA}, \texttt{GEN}, \texttt{QA}, and \texttt{EXIT}), coupled with a prompt bank (see \cref{state machine}), enables the LLM to behave correspondingly across different functions
% \malian{what does role-specific mean?}
, balancing flexibility with control.

The simulation backend evaluates system performance using gem5. The current incarnation of \arch supports gem5 simulations in System Call Emulation (SE) mode. Configuration parameters are converted into gem5 execution commands, and results, including performance statistics, power, and area, are returned to the agent. To minimize simulation overhead, the DSDB supplies precomputed results, significantly speeding up exploration.

A Streamlit-based web UI~\cite{streamlit} supports system setup, simulation control, and visualization. Its interactive features (such as dynamic plotting and public URL access) facilitate remote use and enhance overall accessibility.

% ---------------------------------------------------------------------------------- %
\subsection{DSE AI agent}
The AI agent powers the system, executes the explorations, and analyzes the history parameters and stats. It handles the most complex task of DSE and utilizes the full power of LLMs. However, the LLM itself could hallucinate and generate inaccurate output. To solve these potential problems, we applied several techniques to our AI agent.

\subsubsection{Automation mode switching}
We provide a mode selection button that allows the user to choose between automation and manual modes at any time. In this way, the user can decide whether to let the agent run autonomously or to interact with it for a more customized DSE experience. By default, the agent operates in manual mode, under which an input box is displayed. Upon switching to automation mode, the input box is hidden, and the agent begins running automatically using preset prompts. If the user switches back to manual mode, the agent will wait for the current automation process to complete before displaying the input box again.  

The automation prompts are listed in \cref{tab:auto-prompts}. It should be noted that the prompt in the ANA state is executed at the end of that state to transition to the GEN state, as is the case for other automation prompts. We use automation mode for all experiments to eliminate human influence on the results, while keeping manual mode available for interactive exploration.

\begin{table}[htbp]
\centering
\caption{Prompts for automation mode}
\label{tab:auto-prompts}
\begin{tabular}{@{}lp{6.5cm}@{}}
\toprule
\textbf{State} & \textbf{Prompts} \\
\midrule
\texttt{ANA} &
  Based on the analysis, generate a new parameters batch for the mission. \\
\texttt{GEN} &
  Analyze all the simulation results so far (if we have them) and provide insights that would help better design. \\
\texttt{QA} &
  I have no more questions. Please go to ANA state. \\
\texttt{EXIT} &
  Provide the final parameters batch (the best one) (of size 1) for the mission and exit the program. \\
\bottomrule
\end{tabular}
\end{table}

\subsubsection{Concurrent simulations}
\label{concurrent}
Multiple threads can be used to speed up DSE by running multiple gem5 commands concurrently. {\arch} is built to suit this feature. We design the agent to be able to generate multiple configurations and get their simulation results in each \texttt{GEN} state (see \cref{state machine}). We define \texttt{parameters\_set} as a set of parameters that can be used for a simulation. We define \texttt{parameters\_batch}, or \texttt{batch}, as all the \texttt{parameters\_set} that the agent generated in a \texttt{GEN} state. The number of \texttt{parameters\_set} in a \texttt{parameters\_batch} is called the batch size. Accordingly, we define a \texttt{results\_batch} as all the concurrent results that the simulator returns in a \texttt{GEN} state. {\arch} allows up to 20 concurrent simulations.

\subsubsection{State machine}
\label{state machine}
A state machine structures the LLM’s task execution through a clear, state-driven process. Each state corresponds to one response and is bound to a structured output. The system prompt specifies the initial state, all possible states, and the transitions between them. The LLM’s role in each state is provided in \cref{tab:agent-states}.

\begin{table}[htbp]
\centering
\caption{Agent States and Their Responsibilities}
\label{tab:agent-states}
\begin{tabular}{@{}ll@{}}
\toprule
\textbf{State} & \textbf{Description} \\
\midrule
\texttt{ANA} &
  \begin{tabular}[t]{@{}l@{}}
  - Retrospect past simulation results from chat history. \\
  - Provide CoT-based step-wise analysis and conclusions. \\
  - Explain reasoning and suggest next state (\texttt{ANA} or \texttt{GEN}).
  \end{tabular} \\
\addlinespace
\texttt{GEN} &
  \begin{tabular}[t]{@{}l@{}}
  - Generate a \texttt{parameters\_batch} based on analysis. \\
  - Each \texttt{parameters\_set} respects design space ranges. \\
  - Batch size fits concurrent simulation limit. \\
  - Suggest next state (\texttt{ANA} or \texttt{GEN}).
  \end{tabular} \\
\addlinespace
\texttt{QA} &
  \begin{tabular}[t]{@{}l@{}}
  - Answer user's questions using history and known context. \\
  - Transition back to \texttt{ANA} after answering.
  \end{tabular} \\
\addlinespace
\texttt{EXIT} &
  \begin{tabular}[t]{@{}l@{}}
  - Output final best \texttt{parameters\_batch} (of size 1). \\
  - Triggered when no further improvement is expected.
  \end{tabular} \\
\bottomrule
\end{tabular}
\end{table}

% Strong and reasonable system prompts are essential for the LLM to run the state machine smoothly. \cref{tab:prompt-rules} summarized the prompt generation constraints used in the system prompt under different aspects.

% \begin{table}[htbp]
% \centering
% \caption{Prompt for Agent Generation Constraints}
% \label{tab:prompt-rules}
% \begin{tabular}{@{}lp{4.5cm}@{}}
% \toprule
% \textbf{Aspect} & \textbf{Guidelines} \\
% \midrule
% Design Space Awareness &
%   Agent must obey parameter ranges and design constraints explicitly defined by user. \\
% Concurrent Simulations &
%   Agent should match \texttt{batch} size
%   % \malian{you've used "batch" twice until this point without proper definition}
%   to concurrent simulation capacity. \\
% Output Format &
%   Each \texttt{parameters\_set} must be a Python dictionary (\texttt{dict}) and \texttt{batch} is a list of dictionaries (\texttt{list[dict]}). \\
%   % \malian{is this the definition of a "batch"? If so, it's too late. Define it when you use it for the first time.}. \\
% Transitions &
%   State transitions are determined by current simulation history and user queries. \\
% EXIT Criteria &
%   Agent enters \texttt{EXIT} when no further meaningful exploration is possible. \\
%   % \malian{how about the maximum simulation count parameter?}. \\
% \bottomrule
% \end{tabular}
% \end{table}

\subsubsection{LLM structured output}
A major challenge in using LLMs for parameter generation is their occasional failure to adhere to required formats or content (even with explicit prompting) as observed in earlier models like GPT-3. This is particularly problematic for DSE, which demands interpretable and highly stable parameter outputs. Additionally, the system must output the next state and the LLM’s reasoning behind its parameter choices.

Modern LLMs such as GPT-4 and GPT-4o address this issue through structured output~\cite{schick2023toolformer}, which enables the model’s response to be directly parsed into a Python dictionary. This approach not only ensures consistent and interpretable responses but also reduces output length, thereby lowering generation costs.

\subsubsection{Results Retrospection}
\label{RR}
Results retrospection (RR) is a technique used in our LLM prompting: in every \texttt{ANA} state, instead of directly asking the agent to provide an analysis on history results, which were stored in history chat, we add a "retrospection" prompt:
\begin{quote}
"You should first retrospect on all history simulation results. They are in the chat history, you should read them and print them in batches in a timely manner."
\end{quote}
to let the LLM retrospect on historical results before providing the analysis to prevent LLM from forgetting historical information as the chat length gets longer.

\subsubsection{Baseline Preservation}
\label{BP}
Baseline Preservation (BP) means that in every \texttt{GEN} state, instead of asking the LLM to directly generate a \texttt{batch} without any order, an additional prompt:
\begin{quote}
"When the number of concurrent simulations is larger than 1, you should make the first parameters\_set the best one you can figure out, and use other ones for exploration."
\end{quote}
who asks the AI to preserve the possible best \texttt{parameters\_set} in the first thread while keeping other threads for wild exploring is applied to prevent chaos in the DSE process.

\subsubsection{Chain of Thought Reasoning}
A major challenge in employing LLMs for design space exploration (DSE) is ensuring logically sound and accurate outputs. Chain of Thought (CoT) prompting~\cite{wei2023cot} addresses this by guiding the model to reason step-by-step, improving deliberation over one-shot responses.

In DSE, CoT enables the LLM to systematically evaluate trade-offs. For example, it can reason that increasing L2 cache size typically raises area and power costs, and should be avoided if performance gains are marginal. Such structured reasoning is essential for making informed parameter selections within constrained design spaces.

\subsubsection{Streamlit User-Interface}
\begin{figure}
    \centering
    \includegraphics[width=1\linewidth]{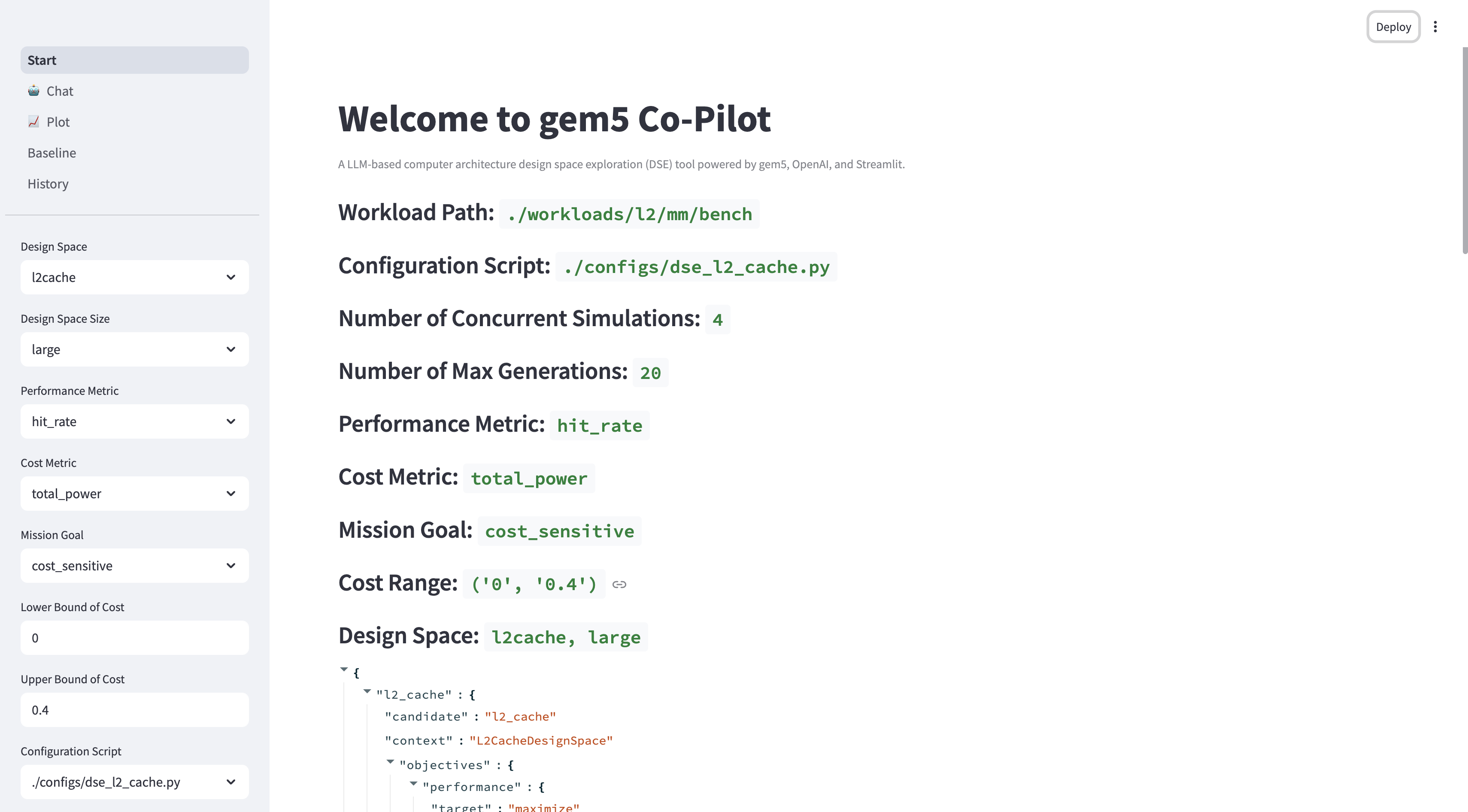}
    \caption{Streamlit UI}
    \label{fig:streamlit}
\end{figure}

We use a popular framework used by today’s web-based AI Applications, Streamlit~\cite{streamlit}, to build our UI, as shown in \cref{fig:streamlit}. It is powerful for displaying chat-bots, DataFrames, dynamic 2D and 3D graphs, curve line analysis, and so on. It is suitable for our DSE setup, interaction, analysis, and result display.

% \begin{figure}
%     \centering
%     \includegraphics[width=1\linewidth]{figs/streamlit_backend.png}
%     \caption{Streamlit hierarchy design}
%     \label{fig:Streamlit hierarchy design}
% \end{figure}

% The Streamlit Page are codes that control the presentation and logic of the UI pages, while the Streamlit Session States codes are built-in Streamlit variable classes that store the data the pages needed. The backend end of the program is integrated into a system.py file, which encapsulates the \textit{agent.py} file that contains the agent powering logic, the \textit{database.py} file that stores the chat and simulation history, and the \textit{simulator.py} file that connects to the gem5 simulator (and the DSDB), as shown in \cref{fig:Streamlit hierarchy design}. We use this design because we want everything in the Streamlit code to be only connected to a single port, here,  \textit{system.py}. This is efficient as we hide all the complex agent logic, database analysis code, and simulator settings behind the system. They provide methods for the system, and system package and use them to interact with Streamlit session states and pages. This design methods make our program easy to test, modify, and extend.

% ---------------------------------------------------------------------------------- %
% \subsection{Simulation Unit}
% The simulation system consists of the gem5 simulator and the the Design Space Description Language (DSDL). We use the gem5 simulator in SE (System Call Emulation) mode. 

\subsection{The gem5 Simulator}

The gem5 simulator is a modular, open-source framework widely used for computer architecture DSE. It provides a flexible tool for simulating CPUs, memory hierarchies, interconnects, and devices.

In our framework, gem5 executes binaries under various configurations generated by the AI agent, producing detailed runtime statistics and performance counters. These outputs are passed to McPAT~\cite{li2009mcpat} for joint power and area estimation, enabling comprehensive design point characterization within constrained optimization objectives.

\subsection{DSDB: A Design Space Database for gem5}
\label{DSDB}

To keep track of the design space exploration, \arch introduces a Design Space Database (DSDB), which is a structured repository of comprehensive simulation results from previously executed runs. The first important use case of DSDB is to establish baseline Pareto-optimal frontiers for DSE. \arch also leverages the DSDB to estimate the performance of a future simulated system without executing a full simulation, significantly accelerating the DSE process.  

In our evaluation, the DSDB provides precomputed Pareto frontiers for benchmarking, enabling objective assessment of \arch’s optimization performance. It also accelerates the \arch loop by retrieving prior results, substantially reducing exploration time and promoting convergence. Although results are fetched from pre-run simulations, each entry corresponds to an actual simulation, ensuring rigor while maintaining efficiency. The simulations populating the DSDB were conducted on a SLURM-managed cluster, as specified in \cref{tab:sim-host}.

%Exploring the vast configuration spaces of modern architectures using cycle-accurate simulators like gem5 is computationally prohibitive. To overcome this, we employ a Design Space Database (DSDB)—a structured repository of comprehensive simulation results across diverse design configurations. The DSDB serves two key roles: it establishes baseline Pareto-optimal frontiers for full design spaces, and acts as a simulation “emulator” to accelerate iterative experimentation within \arch.

%In our evaluation, the DSDB provides precomputed Pareto frontiers for benchmarking, enabling objective assessment of Co-Pilot’s optimization performance. It also speeds up the Co-Pilot loop by retrieving prior results, substantially cutting exploration time and promoting convergence. Although results are fetched from pre-run simulations, each entry corresponds to an actual simulation, ensuring rigor while maintaining efficiency. The simulations populating the DSDB were conducted on a SLUR-managed cluster, as specified in \cref{tab:sim-host}.

\section{Methodology}

\begin{table}[htbp]
    \centering
    \caption{Simulation Environment Configuration}
    \label{tab:sim-host}
    \begin{tabular}{@{}lp{5cm}@{}}
        \toprule
        \textbf{Field} & \textbf{Specification} \\
        \midrule
        \textbf{CPU Model} & Intel Xeon E5-2660 @ 2.20\,GHz (per node) \\
        \textbf{Cores per Job} & 1 core per SLURM task \\
        \textbf{Memory per Job} & 4\,GiB DRAM \\
        \textbf{Operating System} & Rocky Linux 9.4 \\
        \textbf{Execution Model} & Cluster-parallel: one simulation per SLURM job \\
        \bottomrule
    \end{tabular}
\end{table}

\subsection{L2 Cache Design Space Specification}

\cref{tab:l2-large-design-space} defines the large L2 cache design space used in all subsequent experiments. This space spans five key microarchitectural parameters: cache size, associativity, number of MSHRs, MSHR target count, and replacement policy. Each parameter is assigned a broad and representative range of values, capturing realistic design options found in modern systems. When all combinations are enumerated, the design space contains over 7,770 unique configurations, enabling thorough exploration of architectural trade-offs across performance, power, and area.

\begin{table}[htbp]
    \centering
    \caption{L2 Cache Design Space (Large)}
    \label{tab:l2-large-design-space}
    \begin{tabular}{@{}lp{5cm}@{}}
        \toprule
        \textbf{Parameter} & \textbf{Values} \\
        \midrule
        \texttt{l2\_size} &
          128KiB, 256KiB, 512KiB, 1MiB, 2MiB \\
        \texttt{l2\_assoc} &
          2, 4, 8, 16, 32 \\
        \texttt{l2\_mshrs} &
          16, 32, 64 \\
        \texttt{l2\_mshr\_tgts} &
          6, 8, 10, 12, 14, 16, 18, 20, 22, 24, 26, 28, 30, 32 \\
        \texttt{l2\_policy} &
          LRURP, RandomRP, FIFO, FIFORP, LFURP, BIPRP, MRURP, BRRIPRP, SecondChanceRP \\
        \bottomrule
    \end{tabular}
\end{table}

\subsection{DSE Goal and Metrics}
Architectural design is all about trade-offs. We set our system goal to a function of performance and cost \texttt{GOAL = f(PERF, COST)}. The performance 
% \malian{is "PERF" short of "performance"? If so please clearly define.} 
and cost are two entangled variables which in a DSE problem architects must take into consideration.
% \malian{this sentence is grammatically wrong. Please use AI tools to fix your grammar. }. 
In our study, we set our \texttt{GOAL} to be finding the design point that achieves the highest performance in a given cost range.
% \texttt{PP-COST}\malian{fix grammar}. 
We call this point the performance-under-constraint point (\texttt{P\textsubscript{Constraint}P}) and find that it is both DSE-representative and LLM-friendly, due to the concept being easily understood by the model. There may be more than one \texttt{P\textsubscript{Constraint}P} in a given cost range and we regard them all as valid and reasonable goals.

To evaluate the effectiveness of \arch, we use \texttt{l2\_hit\_rate} as the performance (\texttt{PERF}) metric. The \texttt{l2\_hit\_rate} is provided by gem5 stats and is calculated in Equation~\ref{eq:perf}.
%We choose \texttt{l2\_hit\_rate} to be our \texttt{PERF} in our system. \texttt{l2\_hit\_rate} is provided by gem5 stats and calculated in Equation~\ref{eq:perf}.

\begin{align}
    PERF_{\text{hit\_rate}} &= 1 - R_{\text{miss}}
    \label{eq:perf}
\end{align}
{\footnotesize
\begin{align*}
    R_{\text{miss}}  &= \texttt{stats.l2cache.overallMissRate.total}
\end{align*}
}
We choose \texttt{total\_power} as our \texttt{COST}. It is provided by the McPAT 
% \malian{be consistent in spelling McPAT in the paper. Earlier you spelled it as "McPAT"}
power modeling tool and calculated in Equation~\ref{eq:cost}.

\begin{align}
    COST_{\text{total\_power}} &= P_{\text{dyn}} + P_{\text{gate}} + P_{\text{sub}}
    \label{eq:cost}
\end{align}
{\footnotesize
\begin{align*}
    P_{\text{dyn}}  &= \texttt{mcpat.total\_l2s.runtime\_dynamic} \\
    P_{\text{gate}} &= \texttt{mcpat.total\_l2s.gate\_leakage} \\
    P_{\text{sub}}  &= \texttt{mcpat.total\_l2s.subthreshold\_leakage}
\end{align*}
}

We use \texttt{(perf\_ratio, nsims)} pairs as the metric for our goal. We define \texttt{perf\_ratio} as the proportion of a DSE agent generated \texttt{PERF} result value to the \texttt{PERF} value of \texttt{P\textsubscript{Constraint}P} in a given range of \texttt{COST}. 
%\malian{I don't follow this sentence}
To be more specific, when we do DSE within \texttt{COST} range [0,0.15]
%\malian{what does a COST rage mean? Does it mean that you want  to have (in this case) the power to to be under 0.15 Watts? If so, then why do you have a lower bound for the cost?}
%\zuomin{yes, it means the power to be under 0.15 Watts. We have the lower range because we want to represent this "cost constraint" in intervel format. All the cost ranges start with 0, so it only has a higher bound, as in \cref{tab:ppc-results} }
, if the DSE agent finally finds a potential best design point that achieves a \texttt{PERF} value of $P_{DSE}$, while the ground truth best \texttt{PERF} value of \texttt{P\textsubscript{Constraint}P} is $P_{\texttt{P\textsubscript{Constraint}P}}$, then:
\begin{center}
    $\texttt{perf\_ratio}_{[0,0.15]} = \dfrac{P_{DSE}}{P_{\texttt{P\textsubscript{Constraint}P}}}$
\end{center}
In our experiment, a \texttt{perf\_ratio} higher than 97\% is regarded as a realistic design point.

\texttt{nsims} is the number of simulations the agent applies before a DSE achieves its final result. The number of GEN stages used is also recorded as it measures the speed of convergence. The relationship between the number of GEN stages and \texttt{nsims} can be shown as:

\begin{center}
    $\texttt{nsims}=nGENs\times concurrent\_sims\_per\_GEN$
\end{center}

Overall, the \texttt{(perf\_ratio, nsims)} pairs measure a system's ability to achieve a good result while considering the simulation cost.

\subsection{Experiment Design}
\label{exp-design}

We evaluate our approach using a series of cost ranges: \texttt{[0,0.12], [0,0.15], [0,0.2], [0,0.4]}, which represent distinct regions of the cost-performance Pareto front. For each range, we conduct DSE using both baseline methods and \arch, recording the achieved \texttt{perf\_ratio} and number of simulations (\texttt{nsims}).

We compare against two baseline methods: Random Search (RS) and a Genetic Algorithm (GEN)~\cite{mehri2016genetic_fpga}. For RS, we report the first configuration achieving \texttt{perf\_ratio} >97\% and its simulation cost. The GEN implementation uses a population of 50 over 20 generations, with 80\% crossover and 20\% mutation rates, employing tournament selection and elitism. All candidates undergo dynamic validity checks against the \texttt{P\textsubscript{Constraint}P} objective.

We further examine the impact of concurrent simulation count and key prompt strategies (Results Retrospection-RR and Baseline Preservation-BP). Each configuration was run three times, with the best result selected to account for LLM randomness. 
%\malian{why do you run 3 times?}.
%\zuomin{because the results are unstable due to the trait of LLM API. Unrealistic result would happen if we run only once, although with little chance. 3 times running would ensure (not 100\% but high, I assume around 90\%. This is hard to quantize) a better result compared to baseline models.}

A small number of runs failed due to rare LLM output irregularities (e.g., malformed Python dictionaries). These were automatically detected and corrected by requesting a renewed response, ensuring continuous operation.

All simulations use the gem5 System Call Emulator (SE) mode. The baseline configuration (\cref{tab:baseline-config}) features an O3 CPU, two-level caches, and DDR3 memory, with several L2 parameters available for DSE tuning.

The evaluation workload is a blocked matrix multiplication (128$\times$128 matrices, block size 32) in C. With a footprint of $\sim$393 KB (exceeding L2 capacity), it stresses the memory hierarchy while maintaining short runtimes. %We add print functions for the result matrix elements to prevent compiler optimization.
%\malian{I don't follow this sentence!}
%\zuomin{I think this is a trick for writing C code benchmark. Without printf() sometimes the compiler would be smart enough to optimize the code. I don't know how the compiler do this, we use this more as a convention. The Design Space plot looks fine so I think this is not that important.}

Experiments were conducted on both a SLURM cluster and a high-end workstation (Intel i9-13900HX, 64GB RAM). The full design space (7,770 configurations) completes in $\sim$6 hours on the cluster. On the workstation, a single simulation takes $\sim$1 minute, and a full DSE (20 generations) finishes in $\sim$30 minutes.

The OpenAI API cost for a full automated DSE run remains below \$0.50 USD, compared to over \$8 for even a brief manual DSE (10-minute) under a \$50 per hour labor cost
% \malian{what is "manual evaluation at a \$50/hour rate"?}
, demonstrating significant cost efficiency.

\begin{table}[htbp]
    \caption{Baseline gem5 System Configuration}
    \label{tab:baseline-config}
    \centering
    \begin{tabular}{@{}lp{3.6cm}@{}}
        \toprule
        \textbf{Component} & \textbf{Configuration} \\
        \midrule
        ISA & X86 \\
        CPU model & O3 (out-of-order) \\
        Clock frequency & 1~GHz \\
        Memory mode & Timing \\
        Main memory size & 512~MiB \\
        \addlinespace
        L1 I-cache & 16~KiB, 2-way, latency: 2 cycles \\
        L1 D-cache & 64~KiB, 2-way, latency: 2 cycles \\
        L2 cache & 256~KiB, 8-way, latency: 20 cycles (DSE tunable) \\
        L2 MSHRs / Targets per MSHR & 20 / 12 (DSE tuneable) \\
        L2 Replacement policy & LRU (DSE tuneable) \\
        \addlinespace
        System buses & L2XBar and SystemXBar \\
        Memory type & DDR3\_1600\_8x8 \\
        \bottomrule
    \end{tabular}
\end{table}

\section{Results}

\subsection{Performance-Cost Optimization}
Our experimental evaluation demonstrates the effectiveness of \arch in navigating the architectural design space compared to traditional approaches. The results highlight both the efficiency and optimization capabilities of \arch across different cost constraints.

\begin{figure*}[t]
\centering
\begin{subfigure}[t]{0.48\textwidth}
\centering
\includegraphics[width=\linewidth]{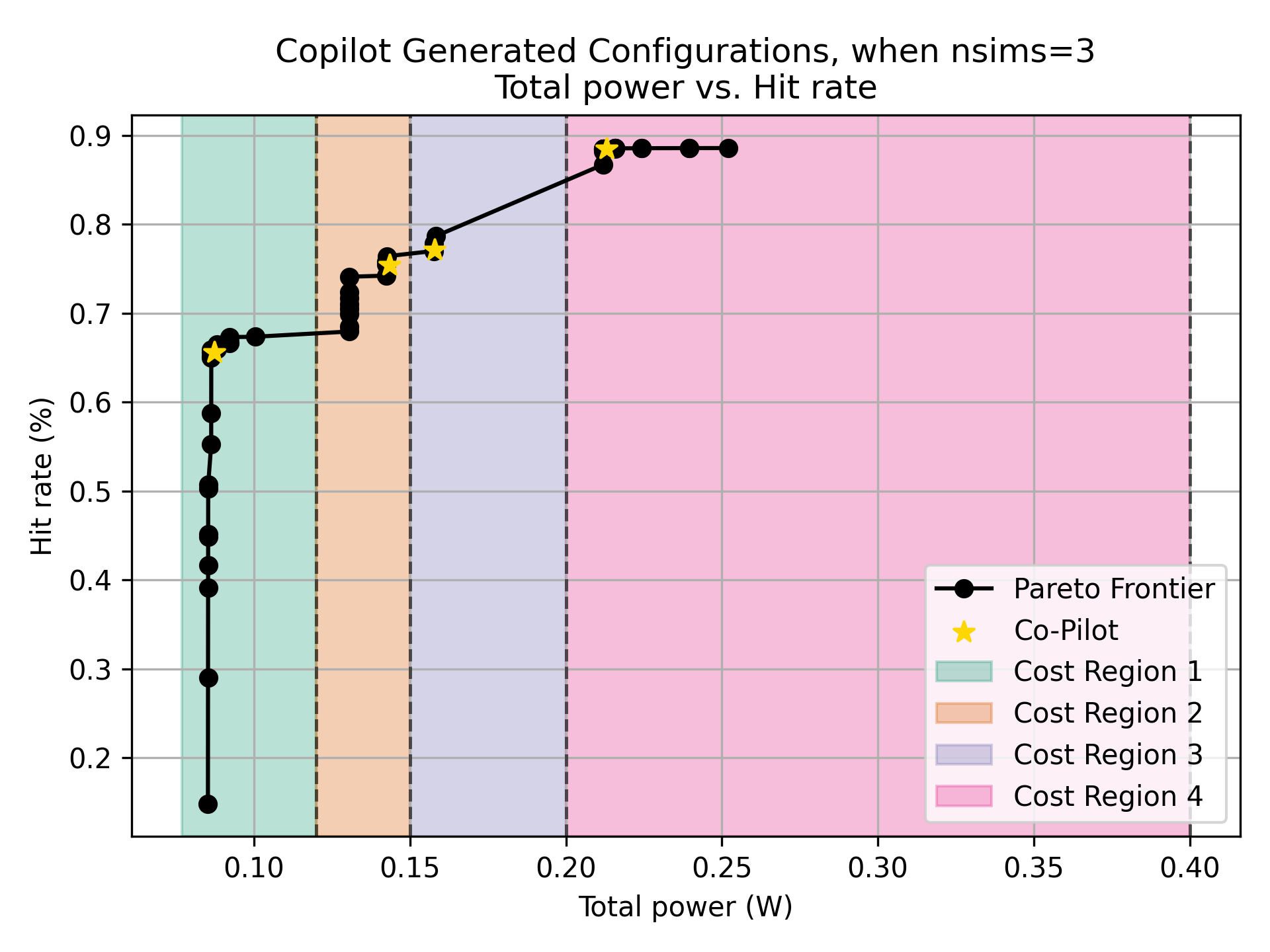}
\caption{3 Concurrent Sims}
\end{subfigure}
\hfill
\begin{subfigure}[t]{0.48\textwidth}
\centering
\includegraphics[width=\linewidth]{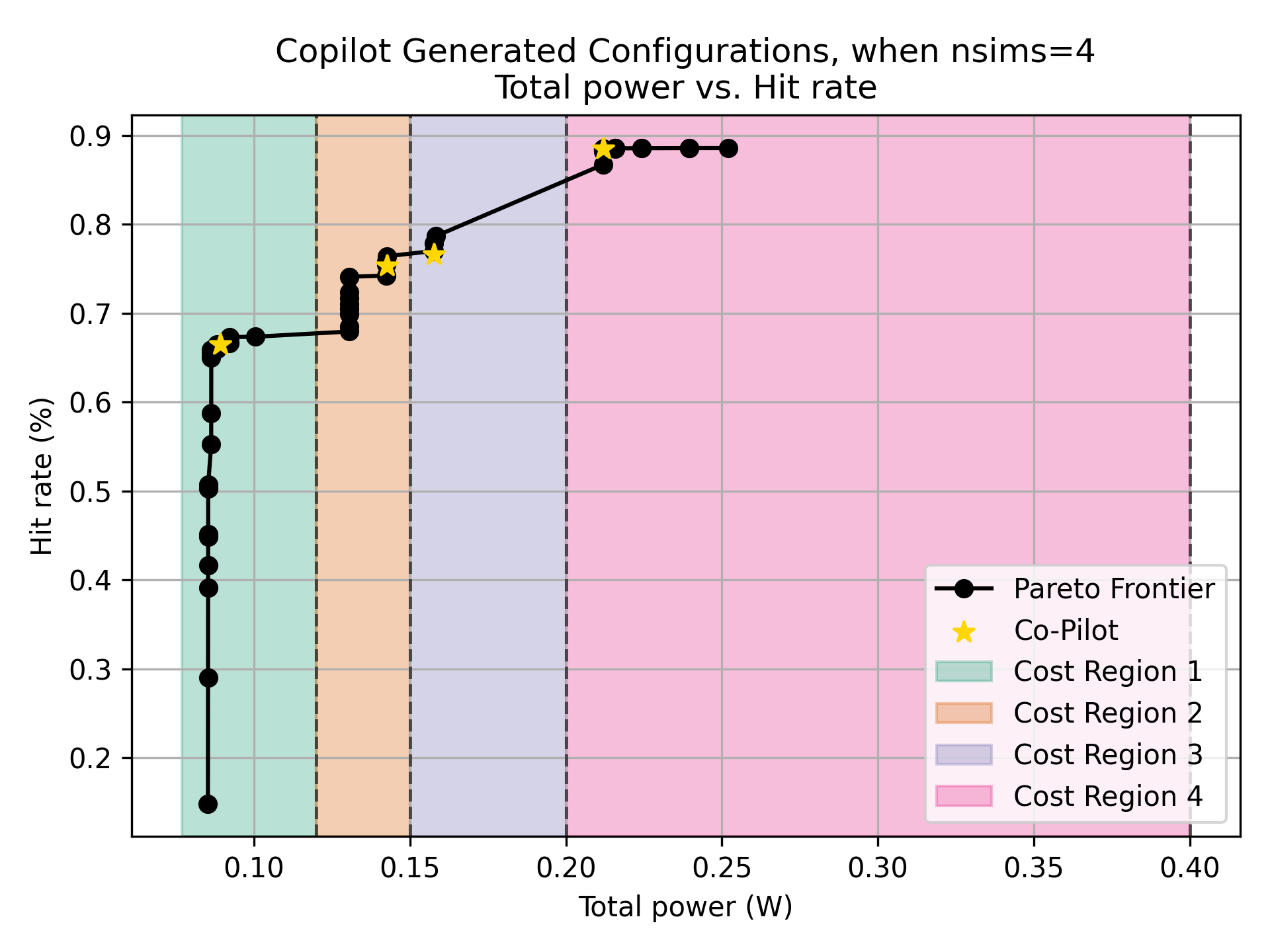}
\caption{4 Concurrent Sims}
\end{subfigure}

\caption{Co-Pilot generated points (yellow) when optimizing total power vs. hit rate, superimposed on the full Pareto frontier (black) under 3 and 4 concurrent simulations. The Pareto frontier is extracted from the DSDB dataset. 
% using a skyline algorithm
%\malian{how do you populate DSDB? Don't you need to run these simulations?}\zuomin{We run the entire Design Space so we have the entire DSDB. we do this to save time on experimenting from running things repetitively. for users who has an unknown design space and not able to run the entire design space, DSDB is initially empty, but would gather simulation results as they use \arch. this would save time if user's future DSEs needs to run former experiments.}
% \malian{what is a skyline algorithm? Add a discription with propser citation in the body of the paper for this.}
The cost thresholds—[0,~0.12], [0,~0.15], [0,~0.2], and [0,~0.4]—represent different ascending stages along the frontier. The experiment system uses an O3 CPU with 2-level caches running a blocked matrix multiplication benchmark (see \cref{exp-design}).}
\label{fig:copilot-power-hitrate}
\end{figure*}

\Cref{fig:copilot-power-hitrate} presents the Pareto-optimal frontier for the full design space with Co-Pilot's selected design points overlaid at 3 and 4 concurrent simulations. These configurations demonstrate \arch's ability to effectively navigate the performance-cost trade-off space while utilizing different levels of parallel simulation resources. The selected points cluster near the Pareto frontier across all cost ranges, indicating effective optimization under varying constraint levels.
%\malian{In figure 4, do you exhaustively run many simulations to identify the Pareto frontier points? }
%\zuomin{We ran all possible design points to get the whole picture of the entire design space, and then use the skyline algorithm to attain the ground-truth frontier. we then generate the optimal design point in a given cost range with \arch and finds that generated points are on the frontier. we do this in our paper because we want to present that \arch can indeed find the optimal design point. Otherwise, we won't be able to know the effectiveness of our system. However, users don't need to do this because our system has been proved to be able to find frontier points in this paper.}

Note that we evaluated all possible design points to obtain a complete view of the design space and then applied the skyline algorithm to determine the ground-truth Pareto frontier. We subsequently used \arch to generate the optimal design point within a given cost range and observed that the generated points lie on the frontier. We perform this exhaustive evaluation in our paper to demonstrate that \arch can indeed identify optimal design points. In practice, however, users do not need to repeat this process, as our results have already validated \arch’s ability to discover frontier points effectively.

% TABLE SUMMARY
\begin{table}[htbp]
    \caption{Performance-Cost Optimization Results (Performance ratio and simulation utilization)}
    \label{tab:ppc-results}
    \centering
    \begin{tabular}{@{}lcccc@{}}
        \toprule
        Model & \multicolumn{4}{c}{Cost Range (W)} \\
        \cmidrule(lr){2-5}
         & [0,0.12] & [0,0.15] & [0,0.2] & [0,0.4] \\
        \midrule
        RS & 0.978 (25) & 0.986 (11) & 0.980 (22) & 0.997 (6) \\
        GEN & 1.000 (133) & 1.000 (158) & 1.000 (137) & 1.000 (127) \\
        \addlinespace
        C-P-1 & 0.980 (6×1) & 0.987 (5×1) & 0.989 (8×1) & 0.999 (2×1) \\
        C-P-2 & 0.978 (5×2) & 0.992 (6×2) & 0.989 (6×2) & 0.998 (1×2) \\
        C-P-3 & 0.974 (4×3) & 0.987 (3×3) & 0.980 (3×3) & 1.000 (1×3) \\
        C-P-4 & 0.987 (2×4) & 0.986 (3×4) & 0.973 (1×4) & 1.000 (1×4) \\
        \bottomrule
    \end{tabular}
    
    \smallskip
    \footnotesize
    \textit{Note:} Values show \texttt{perf\_ratio} achievement (higher is better) and required simulations ($nsims=nGENs\times concurrent\_sims\_per\_GEN$, in parentheses). 
    RS = Random Search baseline; GEN = Genetic Algorithm baseline, C-P-i = Co-Pilot with i concurrent simulations (1-4). 
    All Co-Pilot variants use the DSDB framework.
\end{table}

\Cref{tab:ppc-results} compares the performance ratios and simulation counts across different methods and cost ranges. The Genetic Algorithm (GEN) achieves perfect optimization (1.0) but requires substantially more simulations (127–158). Random Search (RS) attains high performance (0.978–0.997) with moderate simulation counts (6–25), though with greater variability.

Co-Pilot demonstrates superior efficiency, achieving near-optimal performance (0.973–1.000) with significantly fewer simulations (as low as 1–8). Key observations include:
\begin{itemize}
\item \textbf{Rapid Convergence}: Co-Pilot reaches >97\% of optimal within 1–8 simulations, vastly outperforming RS (11–25) and GEN (127–158). For example, in the tight [0,0.12] range, C-P-4 attains 98.7\% with only 2 simulations.
\item \textbf{Scaling with Concurrency}: Higher concurrency (e.g., C-P-3, C-P-4) proves that \arch performs well under different concurrences and total simulation count decreases consistently as concurrency increases, %\malian{what do you mean by "consistency"?}\zuomin{It means first \arch performs well regardless of different concurrency, second the total simulation count decreases consistently.}
especially under cost range [0,0.12], [0,0.15], and [0,0.02].
\item \textbf{Constraint Sensitivity}: All methods perform best in the widest cost range [0,0.4], where Co-Pilot achieves nearly perfect results in just 1–2 simulations.
\end{itemize}

\subsection{Ablation Study}
\label{sec:ablation}
We conduct an ablation study to quantify the contribution of individual components in \arch, focusing on key prompt engineering elements and their impact on design space exploration.

Two critical prompt components are evaluated through removal experiments, where each is systematically omitted and the resulting system is compared against the full configuration:

\begin{itemize}
\item \textbf{Results Retrospection (RR)}: Removes the result retrospection (see \cref{RR}) in the \texttt{ANA} state. The agent does not review historical simulation results.
\item \textbf{Baseline Preservation (BP)}: Removes baseline preservation (see \cref{BP}) in the \texttt{GEN} state. The agent does not retain the best-performing \texttt{parameters\_set} in the initial thread.
\end{itemize}

\begin{table}[htbp]
\caption{Ablation Study Results (\texttt{perf\_ratio} degradation) under Co-Pilot 3-sim}
\label{tab:ablation}
\centering
\begin{tabular}{@{}lcccc@{}}
\toprule
Ablation & \multicolumn{4}{c}{Cost Range (mW)} \\
\cmidrule(lr){2-5}
 & [0,0.12] & [0,0.15] & [0,0.2] & [0,0.4] \\
\midrule
Full System & 0.974 (3) & 0.987 (3) & 0.980 (3) & 1.000 (1) \\
No RR & 0.000 ($\infty$) & 0.970 (1) & 0.944 (4) & 1.000 (1) \\
No BP & 0.974 (6) & 0.987 (8) & 0.959 (8) & 1.000 (2) \\
No RR+No BP & 0.974 (13) & 0.972 (3) & 0.942 (5) & 1.000 (1) \\
\bottomrule
\end{tabular}

\footnotesize
\textit{Note:} Values represent (perf\_ratio achieved / GEN count). Simulation per GEN is fixed at 3.
RR = Results Retrospection (historical simulation analysis), 
BP = Baseline Preservation strategy. 
"$\infty$" indicates failure to converge within 20 simulation cycles. 
All experiments use the same initial seed configuration. 
The [0,0.12] range shows highest sensitivity to component ablation.
\end{table}

Table~\ref{tab:ablation} summarizes the performance impact of ablating key prompt components across cost ranges. Results show that Results Retrospection (RR) is critical under tight constraints: removing RR alone in the [0,0.12] mW range causes complete failure (\texttt{perf\_ratio} = 0),
while also degrading performance in [0,0.15] and [0,0.2] ranges. Baseline Preservation (BP) mainly improves efficiency—without BP, the number of generations increases by 100–167\% across most ranges, though final 
\texttt{perf\_ratio} remains high ($\geq$97.4\% in three ranges). Their 
interaction is complementary: BP partially compensates for RR's absence in 
strict constraints, maintaining performance at the cost of more simulations.
In relaxed constraints ([0,0.4] mW), both components become unnecessary, as all configurations achieve \texttt{perf\_ratio} = 1.0 within 1–2 simulations.

\section{Related Work}
Recent studies have explored using large language models (LLMs) in hardware design and design space exploration (DSE). For high-level synthesis (HLS), LLM-DSE~\cite{llm-dse} proposed a multi-agent system that optimizes compiler directives through specialized roles. Although effective, such approaches remain at a high abstraction level and depend heavily on synthesis feedback from tools like Merlin and Vitis.

Other works, such as CodeChain~\cite{llm-codegen}, generate Verilog or HLS code from natural language prompts. These typically perform one-time synthesis with minimal iteration and often omit integration with performance or power evaluation.

In comparison, our work targets cycle-accurate architectural exploration. The gem5 Co-Pilot embeds LLM reasoning within a state-managed exploration loop that directly manipulates hardware parameters. Each design is assessed using gem5 for performance and McPAT for power and area, creating a closed-loop refinement process based on simulation feedback. Unlike scripted methods~\cite{gem5-mcpat}, Co-Pilot employs natural language planning to interpret trade-offs and navigate the design space efficiently.

This approach aligns with the agentic, cross-layer vision of Architecture 2.0~\cite{arch2_0}, where AI agents use system feedback to guide optimization. Co-Pilot implements this through declarative design specification, intelligent control, and cycle-level evaluation, bridging high-level intent and low-level implementation in architectural DSE.

\section{Conclusion}
In this work, we introduced \arch, a generative AI-powered assistant for architectural design space exploration using gem5. We evaluated \arch with a simple yet tractable example of design space exploration for efficient cache design and demonstrated its effectiveness. \arch can significantly reduce the time required to explore the design space of future hardware architectures by intelligently navigating different configurations---avoiding unnecessary simulations whose results can either be predicted (by maintaining a design space database) or ruled out through reasoning. The design of \arch is modular, making it both extensible and flexible for future use not only with gem5 but also with other architectural tools.

%\malian{I will write the conclusion once the rest of the paper is ready}

\begin{acks}

This work was supported in part by NSF award 2519295. We thank Ampere Computing for their generous server donation that we used to run simulations. Any opinions, findings, conclusions, and recommendations expressed in this material are those of the authors and do not necessarily reflect those of the sponsors.

\end{acks}

\bibliographystyle{ACM-Reference-Format}
\bibliography{sections/reference}


\begin{thebibliography}{22}

%%% the Web address.

\ifx \showCODEN    \undefined \def \showCODEN     #1{\unskip}     \fi
\ifx \showDOI      \undefined \def \showDOI       #1{#1}\fi
\ifx \showISBNx    \undefined \def \showISBNx     #1{\unskip}     \fi
\ifx \showISBNxiii \undefined \def \showISBNxiii  #1{\unskip}     \fi
\ifx \showISSN     \undefined \def \showISSN      #1{\unskip}     \fi
\ifx \showLCCN     \undefined \def \showLCCN      #1{\unskip}     \fi
\ifx \shownote     \undefined \def \shownote      #1{#1}          \fi
\ifx \showarticletitle \undefined \def \showarticletitle #1{#1}   \fi
\ifx \showURL      \undefined \def \showURL       {\relax}        \fi
% invisible to TeX
\providecommand\bibfield[2]{#2}
\providecommand\bibinfo[2]{#2}
\providecommand\natexlab[1]{#1}
\providecommand\showeprint[2][]{arXiv:#2}

\bibitem[Bai et~al\mbox{.}(2021)]%
        {Arch-2021ICCAD-BOOM-Explorer}
\bibfield{author}{\bibinfo{person}{Chen Bai}, \bibinfo{person}{Qi Sun}, \bibinfo{person}{Jianwang Zhai}, \bibinfo{person}{Yuzhe Ma}, \bibinfo{person}{Bei Yu}, {and} \bibinfo{person}{Martin~DF Wong}.} \bibinfo{year}{2021}\natexlab{}.
\newblock \showarticletitle{{BOOM-Explorer: RISC-V BOOM Microarchitecture Design Space Exploration Framework}}. In \bibinfo{booktitle}{\emph{2021 IEEE/ACM International Conference on Computer Aided Design (ICCAD)}}. IEEE, \bibinfo{pages}{1--9}.
\newblock


\bibitem[Basile et~al\mbox{.}(2003)]%
        {genetic-dse}
\bibfield{author}{\bibinfo{person}{C. Basile}, \bibinfo{person}{D. Bertozzi}, \bibinfo{person}{A. Jantsch}, {and} \bibinfo{person}{M. Sami}.} \bibinfo{year}{2003}\natexlab{}.
\newblock \showarticletitle{Automatic design space exploration using genetic algorithms}. In \bibinfo{booktitle}{\emph{Design, Automation and Test in Europe Conference and Exhibition}}. IEEE, \bibinfo{pages}{274--279}.
\newblock


\bibitem[Benassi et~al\mbox{.}(2011)]%
        {bayesian-dse}
\bibfield{author}{\bibinfo{person}{R. Benassi}, \bibinfo{person}{J. Bect}, {and} \bibinfo{person}{E. Vazquez}.} \bibinfo{year}{2011}\natexlab{}.
\newblock \showarticletitle{Efficient multi-objective design space exploration using surrogate models}.
\newblock \bibinfo{journal}{\emph{Structural and Multidisciplinary Optimization}} \bibinfo{volume}{43}, \bibinfo{number}{4} (\bibinfo{year}{2011}), \bibinfo{pages}{507--525}.
\newblock


\bibitem[Binkert et~al\mbox{.}(2011)]%
        {binkert2011gem5}
\bibfield{author}{\bibinfo{person}{Nathan Binkert}, \bibinfo{person}{Bradford Beckmann}, \bibinfo{person}{Gabriel Black}, \bibinfo{person}{Steven~K. Reinhardt}, \bibinfo{person}{Ali Saidi}, \bibinfo{person}{Arkaprava Basu}, \bibinfo{person}{Joel Hestness}, \bibinfo{person}{Derek~R. Hower}, \bibinfo{person}{Tushar Krishna}, \bibinfo{person}{Somayeh Sardashti}, \bibinfo{person}{Rathijit Sen}, \bibinfo{person}{Korey Sewell}, \bibinfo{person}{Muhammad Shoaib}, \bibinfo{person}{Nilay Vaish}, \bibinfo{person}{Mark~D. Hill}, {and} \bibinfo{person}{David~A. Wood}.} \bibinfo{year}{2011}\natexlab{}.
\newblock \showarticletitle{The gem5 simulator}.
\newblock \bibinfo{journal}{\emph{SIGARCH Comput. Archit. News}} \bibinfo{volume}{39}, \bibinfo{number}{2} (\bibinfo{date}{Aug.} \bibinfo{year}{2011}), \bibinfo{pages}{1–7}.
\newblock
\showISSN{0163-5964}
\urldef\tempurl%
\url{https://doi.org/10.1145/2024716.2024718}
\showDOI{\tempurl}


\bibitem[Brooks and Martonosi(2001)]%
        {simulated-annealing}
\bibfield{author}{\bibinfo{person}{David Brooks} {and} \bibinfo{person}{Margaret Martonosi}.} \bibinfo{year}{2001}\natexlab{}.
\newblock \showarticletitle{Automated design space exploration using simple analytic models}. In \bibinfo{booktitle}{\emph{Proceedings of the 2001 International Conference on Supercomputing (ICS)}}. ACM, \bibinfo{pages}{24--33}.
\newblock
\urldef\tempurl%
\url{https://doi.org/10.1145/377792.377801}
\showDOI{\tempurl}


\bibitem[Li et~al\mbox{.}(2009)]%
        {li2009mcpat}
\bibfield{author}{\bibinfo{person}{Sheng Li}, \bibinfo{person}{Jung~Ho Ahn}, \bibinfo{person}{Richard~D. Strong}, \bibinfo{person}{Jay~B. Brockman}, \bibinfo{person}{Dean~M. Tullsen}, {and} \bibinfo{person}{Norman~P. Jouppi}.} \bibinfo{year}{2009}\natexlab{}.
\newblock \showarticletitle{McPAT: An integrated power, area, and timing modeling framework for multicore and manycore architectures}. In \bibinfo{booktitle}{\emph{2009 42nd Annual IEEE/ACM International Symposium on Microarchitecture (MICRO)}}. \bibinfo{pages}{469--480}.
\newblock


\bibitem[Li et~al\mbox{.}(2004)]%
        {Li_ICS04}
\bibfield{author}{\bibinfo{person}{Xiaoqiao Li}, \bibinfo{person}{Tulika Mitra}, {and} \bibinfo{person}{Abhik Roychoudhury}.} \bibinfo{year}{2004}\natexlab{}.
\newblock \showarticletitle{Design Space Exploration of Caches Using Compressed Traces}. In \bibinfo{booktitle}{\emph{Proc. 18th Int'l Conf. on Supercomputing (ICS)}}. \bibinfo{pages}{243--252}.
\newblock


\bibitem[Liu et~al\mbox{.}(2021)]%
        {rl-dse}
\bibfield{author}{\bibinfo{person}{Hanrui Liu}, \bibinfo{person}{Haicheng Wei}, \bibinfo{person}{Yujun Guo}, \bibinfo{person}{Jiacheng Wu}, {and} \bibinfo{person}{Vivienne Sze}.} \bibinfo{year}{2021}\natexlab{}.
\newblock \showarticletitle{Hardware architecture search with reinforcement learning}.
\newblock \bibinfo{journal}{\emph{International Conference on Learning Representations (ICLR)}} (\bibinfo{year}{2021}).
\newblock


\bibitem[Lowe-Power et~al\mbox{.}(2020)]%
        {lowe2020gem5}
\bibfield{author}{\bibinfo{person}{Jason Lowe-Power}, \bibinfo{person}{Abdul~Mutaal Ahmad}, \bibinfo{person}{Ayaz Akram}, \bibinfo{person}{Mohammad Alian}, \bibinfo{person}{Rico Amslinger}, \bibinfo{person}{Matteo Andreozzi}, \bibinfo{person}{Adrià Armejach}, \bibinfo{person}{Nils Asmussen}, \bibinfo{person}{Brad Beckmann}, \bibinfo{person}{Srikant Bharadwaj}, \bibinfo{person}{Gabe Black}, \bibinfo{person}{Gedare Bloom}, \bibinfo{person}{Bobby~R. Bruce}, \bibinfo{person}{Daniel~Rodrigues Carvalho}, \bibinfo{person}{Jeronimo Castrillon}, \bibinfo{person}{Lizhong Chen}, \bibinfo{person}{Nicolas Derumigny}, \bibinfo{person}{Stephan Diestelhorst}, \bibinfo{person}{Wendy Elsasser}, \bibinfo{person}{Carlos Escuin}, \bibinfo{person}{Marjan Fariborz}, \bibinfo{person}{Amin Farmahini-Farahani}, \bibinfo{person}{Pouya Fotouhi}, \bibinfo{person}{Ryan Gambord}, \bibinfo{person}{Jayneel Gandhi}, \bibinfo{person}{Dibakar Gope}, \bibinfo{person}{Thomas Grass}, \bibinfo{person}{Anthony Gutierrez}, \bibinfo{person}{Bagus
  Hanindhito}, \bibinfo{person}{Andreas Hansson}, \bibinfo{person}{Swapnil Haria}, \bibinfo{person}{Austin Harris}, \bibinfo{person}{Timothy Hayes}, \bibinfo{person}{Adrian Herrera}, \bibinfo{person}{Matthew Horsnell}, \bibinfo{person}{Syed Ali~Raza Jafri}, \bibinfo{person}{Radhika Jagtap}, \bibinfo{person}{Hanhwi Jang}, \bibinfo{person}{Reiley Jeyapaul}, \bibinfo{person}{Timothy~M. Jones}, \bibinfo{person}{Matthias Jung}, \bibinfo{person}{Subash Kannoth}, \bibinfo{person}{Hamidreza Khaleghzadeh}, \bibinfo{person}{Yuetsu Kodama}, \bibinfo{person}{Tushar Krishna}, \bibinfo{person}{Tommaso Marinelli}, \bibinfo{person}{Christian Menard}, \bibinfo{person}{Andrea Mondelli}, \bibinfo{person}{Miquel Moreto}, \bibinfo{person}{Tiago Mück}, \bibinfo{person}{Omar Naji}, \bibinfo{person}{Krishnendra Nathella}, \bibinfo{person}{Hoa Nguyen}, \bibinfo{person}{Nikos Nikoleris}, \bibinfo{person}{Lena~E. Olson}, \bibinfo{person}{Marc Orr}, \bibinfo{person}{Binh Pham}, \bibinfo{person}{Pablo Prieto}, \bibinfo{person}{Trivikram
  Reddy}, \bibinfo{person}{Alec Roelke}, \bibinfo{person}{Mahyar Samani}, \bibinfo{person}{Andreas Sandberg}, \bibinfo{person}{Javier Setoain}, \bibinfo{person}{Boris Shingarov}, \bibinfo{person}{Matthew~D. Sinclair}, \bibinfo{person}{Tuan Ta}, \bibinfo{person}{Rahul Thakur}, \bibinfo{person}{Giacomo Travaglini}, \bibinfo{person}{Michael Upton}, \bibinfo{person}{Nilay Vaish}, \bibinfo{person}{Ilias Vougioukas}, \bibinfo{person}{William Wang}, \bibinfo{person}{Zhengrong Wang}, \bibinfo{person}{Norbert Wehn}, \bibinfo{person}{Christian Weis}, \bibinfo{person}{David~A. Wood}, \bibinfo{person}{Hongil Yoon}, {and} \bibinfo{person}{Éder F.~Zulian}.} \bibinfo{year}{2020}\natexlab{}.
\newblock \bibinfo{title}{The gem5 Simulator: Version 20.0+}.
\newblock
\newblock
\showeprint[arxiv]{2007.03152}~[cs.AR]
\urldef\tempurl%
\url{https://arxiv.org/abs/2007.03152}
\showURL{%
\tempurl}


\bibitem[Marujo et~al\mbox{.}(2011)]%
        {aco-dse}
\bibfield{author}{\bibinfo{person}{Pedro Marujo}, \bibinfo{person}{Luiz Zschornack}, \bibinfo{person}{Ney Calazans}, {and} \bibinfo{person}{Fernando Moraes}.} \bibinfo{year}{2011}\natexlab{}.
\newblock \showarticletitle{An ant colony optimization approach to design space exploration of network-on-chip architectures}. In \bibinfo{booktitle}{\emph{Proceedings of the 2011 International Conference on Hardware/Software Codesign and System Synthesis}}. ACM, \bibinfo{pages}{313--322}.
\newblock


\bibitem[Mehri and Alizadeh(2017)]%
        {mehri2016genetic_fpga}
\bibfield{author}{\bibinfo{person}{Hossein Mehri} {and} \bibinfo{person}{Bijan Alizadeh}.} \bibinfo{year}{2017}\natexlab{}.
\newblock \showarticletitle{Genetic‑Algorithm‑Based FPGA Architectural Exploration Using Analytical Models}.
\newblock \bibinfo{journal}{\emph{ACM Transactions on Design Automation of Electronic Systems}} \bibinfo{volume}{22}, \bibinfo{number}{1} (\bibinfo{year}{2017}), \bibinfo{pages}{1--17}.
\newblock
\urldef\tempurl%
\url{https://doi.org/10.1145/2939372}
\showDOI{\tempurl}
\newblock
\shownote{Published online September 2 2016}.


\bibitem[Mei et~al\mbox{.}(2021)]%
        {Mei2021ZigZag}
\bibfield{author}{\bibinfo{person}{Linyan Mei}, \bibinfo{person}{Pouya Houshmand}, \bibinfo{person}{Vikram Jain}, \bibinfo{person}{Sebastian Giraldo}, {and} \bibinfo{person}{Marian Verhelst}.} \bibinfo{year}{2021}\natexlab{}.
\newblock \showarticletitle{{ZigZag: Enlarging Joint Architecture-Mapping Design Space Exploration for DNN Accelerators}}.
\newblock \bibinfo{journal}{\emph{IEEE Trans. Comput.}} \bibinfo{volume}{70}, \bibinfo{number}{8} (\bibinfo{year}{2021}), \bibinfo{pages}{1160--1174}.
\newblock
\urldef\tempurl%
\url{https://doi.org/10.1109/TC.2021.3059962}
\showDOI{\tempurl}


\bibitem[Reddi and Yazdanbakhsh(2025)]%
        {arch2_0}
\bibfield{author}{\bibinfo{person}{Vijay~Janapa Reddi} {and} \bibinfo{person}{Amir Yazdanbakhsh}.} \bibinfo{year}{2025}\natexlab{}.
\newblock \showarticletitle{{ Architecture 2.0: Foundations of Artificial Intelligence Agents for Modern Computer System Design }}.
\newblock \bibinfo{journal}{\emph{Computer}} \bibinfo{volume}{58}, \bibinfo{number}{02} (\bibinfo{date}{Feb.} \bibinfo{year}{2025}), \bibinfo{pages}{116--124}.
\newblock
\showISSN{1558-0814}
\urldef\tempurl%
\url{https://doi.org/10.1109/MC.2024.3521641}
\showDOI{\tempurl}


\bibitem[Schick et~al\mbox{.}(2023)]%
        {schick2023toolformer}
\bibfield{author}{\bibinfo{person}{Timo Schick}, \bibinfo{person}{Jane Dwivedi-Yu}, \bibinfo{person}{Roberto Dess\'{\i}}, \bibinfo{person}{Roberta Raileanu}, \bibinfo{person}{Maria Lomeli}, \bibinfo{person}{Eric Hambro}, \bibinfo{person}{Luke Zettlemoyer}, \bibinfo{person}{Nicola Cancedda}, {and} \bibinfo{person}{Thomas Scialom}.} \bibinfo{year}{2023}\natexlab{}.
\newblock \showarticletitle{Toolformer: language models can teach themselves to use tools}. In \bibinfo{booktitle}{\emph{Proceedings of the 37th International Conference on Neural Information Processing Systems}} (New Orleans, LA, USA) \emph{(\bibinfo{series}{NIPS '23})}. \bibinfo{publisher}{Curran Associates Inc.}, \bibinfo{address}{Red Hook, NY, USA}, Article \bibinfo{articleno}{2997}, \bibinfo{numpages}{13}~pages.
\newblock


\bibitem[Tashiro and Oyamada(2016)]%
        {gem5-mcpat}
\bibfield{author}{\bibinfo{person}{Renan Tashiro} {and} \bibinfo{person}{Marcio~Seiji Oyamada}.} \bibinfo{year}{2016}\natexlab{}.
\newblock \showarticletitle{An Environment for Design Space Exploration Using gem5-McPAT}. In \bibinfo{booktitle}{\emph{2016 VI Brazilian Symposium on Computing Systems Engineering (SBESC)}}. \bibinfo{pages}{220--225}.
\newblock
\urldef\tempurl%
\url{https://doi.org/10.1109/SBESC.2016.042}
\showDOI{\tempurl}


\bibitem[Treuille et~al\mbox{.}(2019)]%
        {streamlit}
\bibfield{author}{\bibinfo{person}{Adrien Treuille}, \bibinfo{person}{Amanda Kelly}, \bibinfo{person}{Thiago Teixeira}, {et~al\mbox{.}}} \bibinfo{year}{2019}\natexlab{}.
\newblock \bibinfo{title}{Streamlit: The fastest way to build data apps}.
\newblock \bibinfo{howpublished}{\url{https://streamlit.io}}.
\newblock
\newblock
\shownote{Accessed: 2025-06-28}.


\bibitem[Wang et~al\mbox{.}(2025)]%
        {llm-dse}
\bibfield{author}{\bibinfo{person}{Hanyu Wang}, \bibinfo{person}{Xinrui Wu}, \bibinfo{person}{Zijian Ding}, \bibinfo{person}{Su Zheng}, \bibinfo{person}{Chengyue Wang}, \bibinfo{person}{Tony Nowatzki}, \bibinfo{person}{Yizhou Sun}, {and} \bibinfo{person}{Jason Cong}.} \bibinfo{year}{2025}\natexlab{}.
\newblock \bibinfo{title}{LLM-DSE: Searching Accelerator Parameters with LLM Agents}.
\newblock
\newblock
\showeprint[arxiv]{2505.12188}~[cs.AR]
\urldef\tempurl%
\url{https://arxiv.org/abs/2505.12188}
\showURL{%
\tempurl}


\bibitem[Wei et~al\mbox{.}(2023)]%
        {wei2023cot}
\bibfield{author}{\bibinfo{person}{Jason Wei}, \bibinfo{person}{Xuezhi Wang}, \bibinfo{person}{Dale Schuurmans}, \bibinfo{person}{Maarten Bosma}, \bibinfo{person}{Brian Ichter}, \bibinfo{person}{Fei Xia}, \bibinfo{person}{Ed Chi}, \bibinfo{person}{Quoc Le}, {and} \bibinfo{person}{Denny Zhou}.} \bibinfo{year}{2023}\natexlab{}.
\newblock \bibinfo{title}{Chain-of-Thought Prompting Elicits Reasoning in Large Language Models}.
\newblock
\newblock
\showeprint[arxiv]{2201.11903}~[cs.CL]
\urldef\tempurl%
\url{https://arxiv.org/abs/2201.11903}
\showURL{%
\tempurl}


\bibitem[Wolf et~al\mbox{.}(2005)]%
        {dse-traditional}
\bibfield{author}{\bibinfo{person}{Marilyn Wolf}, \bibinfo{person}{Ahmed~A Jerraya}, {and} \bibinfo{person}{Grant Martin}.} \bibinfo{year}{2005}\natexlab{}.
\newblock \showarticletitle{A survey of design space exploration for hardware systems}.
\newblock \bibinfo{journal}{\emph{ACM Transactions on Design Automation of Electronic Systems (TODAES)}} \bibinfo{volume}{10}, \bibinfo{number}{3} (\bibinfo{year}{2005}), \bibinfo{pages}{273--279}.
\newblock


\bibitem[Zamfirescu-Pereira et~al\mbox{.}(2025)]%
        {llm-codegen}
\bibfield{author}{\bibinfo{person}{J.D. Zamfirescu-Pereira}, \bibinfo{person}{Eunice Jun}, \bibinfo{person}{Michael Terry}, \bibinfo{person}{Qian Yang}, {and} \bibinfo{person}{Bjoern Hartmann}.} \bibinfo{year}{2025}\natexlab{}.
\newblock \showarticletitle{Beyond Code Generation: LLM-supported Exploration of the Program Design Space}. In \bibinfo{booktitle}{\emph{Proceedings of the 2025 CHI Conference on Human Factors in Computing Systems}} \emph{(\bibinfo{series}{CHI ’25})}. \bibinfo{publisher}{ACM}, \bibinfo{pages}{1–17}.
\newblock
\urldef\tempurl%
\url{https://doi.org/10.1145/3706598.3714154}
\showDOI{\tempurl}


\bibitem[Zhai and Cai(2023)]%
        {Zhai2023ParetoActiveLearning}
\bibfield{author}{\bibinfo{person}{Jianwang Zhai} {and} \bibinfo{person}{Yici Cai}.} \bibinfo{year}{2023}\natexlab{}.
\newblock \showarticletitle{{Microarchitecture Design Space Exploration via Pareto-Driven Active Learning}}.
\newblock \bibinfo{journal}{\emph{IEEE Transactions on Very Large Scale Integration (VLSI) Systems}} (\bibinfo{year}{2023}).
\newblock
\urldef\tempurl%
\url{https://doi.org/10.1109/TVLSI.2023.3311620}
\showURL{%
\tempurl}
\newblock
\shownote{Early Access / preprint in 2023}.


\bibitem[Zhao et~al\mbox{.}(2022)]%
        {archgym}
\bibfield{author}{\bibinfo{person}{Xin Zhao}, \bibinfo{person}{Arka Mishra}, \bibinfo{person}{Aditya Bagaria}, \bibinfo{person}{Amirali Ebrahimi}, \bibinfo{person}{Udit Thakker}, \bibinfo{person}{Hanrui Zhang}, \bibinfo{person}{Yuke Zhang}, \bibinfo{person}{Jason Mars}, \bibinfo{person}{Krste Asanović}, \bibinfo{person}{Vivienne Sze}, {et~al\mbox{.}}} \bibinfo{year}{2022}\natexlab{}.
\newblock \showarticletitle{ArchGym: An Open-Source Gym for Machine Learning-Driven Architecture Research}.
\newblock \bibinfo{journal}{\emph{arXiv preprint arXiv:2205.10371}} (\bibinfo{year}{2022}).
\newblock


\end{thebibliography}

\end{document}